\documentstyle[11pt,amssymb]{article}

\def\be{\begin{equation}}
\def\ee{\end{equation}}

\def\be{\begin{equation}}
\def\ee{\end{equation}}
\title{{\hfill\small\tt Mod.\,Phys.\,Lett.\,{\bf A8},\,No.31,\,2989-2999\,
(1993)}\\
{\hfill}\\
A new example of N=2 supersymmetric\\
Landau-Ginzburg theories: the two-ring case}
\author{A.M. Perelomov
\footnote{On leave of absence from Institute of Theoretical and Experimental
Physics, 117 259 Moscow, Russia}\\
{\small\em Max-Planck-Institut~f\"ur~Mathematik,}\\
{\small\em Gottfried-Claren-Strasse 26, D-5300 Bonn 3, Germany }\\
Shi-shyr Roan\\
{\small\em Department of Mathematics, Tsing-Hua University, Taiwan}
}

\date{}
\begin{document}
\maketitle{}

\begin{abstract}\noindent
The new example of $N=2$ supersymmetric Landau-Ginzburg theories is considered
when the critical values of the superpotential $w(x)$ form the regular
two-ring configuration. It is shown that at the deformation, which does
not change the form of this configuration, the vacuum state metric satisfies
the equation of non-Abelian ${2\times 2}$ Toda system.
\end{abstract}

\noindent
In [1] the authors
have considered $N=2$ supersymmetric Landau-Ginzburg theories and have
showed that in many cases the metric for supersymmetric ground states
for special deformations of this metric satisfies the certain system
of PDE's, such  as Toda equations. For further development related to the
theory see [2], [3].

In the note [4] two new examples of such theories were considered.
The purpose of this letter is to give an additional
example of such theories. Namely we will show here
that the two-ring case gives instead of the standard Toda system,
the so-called non-Abelian ${2\times 2}$ Toda system.

{\bf 1.} Let us remind first of all some basic facts from $N=2$ supersymmetric
Landau-Ginzburg theory (for more details see [1]). The basic quantities
here are the chiral fields $\phi_i$, the vacuum state $|0>$ and the
states
\be |j>=\phi_j|0>. \ee
The action of $\phi_j$ on this state gives
\be \phi_i|j>=\phi_i\phi_j|0>=C_{ij}^k\,\phi_k|0>= C_{ij}^k\,|k>.\ee
So the action of the chiral field $\phi_i$ in the subsector of vacuum
states is given by the matrix $(C_i)_j^k=C_{ij}^k $. Analogously, we have
anti-chiral
fields $\phi_{\bar i}$ and the states $|\bar j>$. So we may define two
metric tensors
\be \eta_{ij}=<j|i>\ee
and
\be g_{i\bar j}=<\bar j|i>,\ee
which should satisfy the condition
\be \eta^{-1}g(\eta^{-1}g)^*=1.\ee
The theory is determined by the superpotential $w(x_j)$ which is
holomorphic function of complex variables $x_i$. The superpotential
completely determines the chiral ring
\be {\cal R}={\bf C}[x_i]/\partial _i w\ee
and we may also determine the metric $\eta _{ij}$ by the formula
\be \eta_{ij}=<i|j>=\mbox{Res}_w [\phi _i \phi _j], \ee
where
\be \mbox{Res}_w [\phi ]=\sum _{dw=0} \phi (x)H^{-1}(x);\qquad
H=\mbox{det}\,(\partial _i\partial _j w). \ee
As for the metric $g_{i\bar j}$, then, as was shown in [1], it should
satisfy the zero-curvature conditions
\be \bar \partial _i(g\partial _j\,g^{-1})-[C_j,g(C_i)^+g^{-1}]=0, \ee
\be \partial _iC_j-\partial _jC_i+ [g(\partial _ig^{-1}),C_j]- [g(\partial _j
g^{-1}),C_i]=0. \ee
Note also that the metric $g_{ij}$ should satisfy the ``reality constraint''
\be \eta ^{-1}g(\eta^{-1}g )^*=1. \ee

{\bf 2.} Let us begin to describe the system. As for the superpotential
$w(x)$ we take
\be w(x) = t\left( \frac{x^{2n+1}}{2n+1} + 2c\,\frac{x^{n+1}}{n+1} - x
\right) .\ee
Then
\be w'(x)=t\left( x^{2n} + 2cx^{n} -1\right) .\ee
So the zeros of $w'(x)$ are located on two rings
\be w'(x)=\prod _{j=0}^{n-1} (x-a_{j})\prod _{j=0}^{n-1}(x-b_j),\ee
where
\be a_{j}=a\omega^{j},\quad  b_{j}=b\omega ^{j}\,\varepsilon \ee
\be \omega =\exp \frac{2i\pi }{n},\quad \varepsilon =\exp \frac{i\pi }n,\quad
a<b,\quad  ab=1,\quad  b^{n}-a^{n}=2c.\ee
So the chiral ring take the form
\be {\cal R} ={\bf C}[x]/w'(x)= {\bf C}[x]/(x^{2n}+ 2cx^n - 1),\ee
so that
\be w''(x)= nx^{n-1} \left[ (x^{n}+ b^{n})+ (x^{n}- a^{n})\right] \ee
and
\be w''(a_{j})= \alpha a_{j}^{-1},\qquad w''(b_{j})= \beta b_{j}^{-1}\ee
with
\be \alpha =na^{n}\left( a^{n}+b^{n}\right) ,\qquad \beta =nb^{n}
\left( a^{n}+ b^{n}\right) .\ee
{\bf 3.} Now using (8) we can calculate the metric tensor
\be \eta _{ij}= <i|j>.\ee
We have
\begin{eqnarray}
\mbox{Res}_{w}(x^k)&=& \sum _{dw= 0}\frac{x^k}{w''(x)}=
\sum _{j=0}^{n-1}\left( \frac{a_j^{k+1}}{\alpha}+\frac{b_j^{k+1}}{\beta}
\right) \nonumber \\
&=&\left( \frac{a^{k+1}}{\alpha }+ (-1)^{\frac{k+1}n}\,\frac{b^{k+1}}{\beta}
\right) \sum _{j=0}^{n-1}(\omega ^{j})^{k+1}\nonumber \\
&=&\left\{ \begin{array}{ll}0&\mbox{if}\,\,n\nmid (k+1),\\
\frac{a^{k+1-n}+(-1)^{(k+1)/n}\,b^{k+1-n}}{t(a^{n}+ b^{n})}&\mbox{if}\,\,
n|(k+1).\end{array}\right. \end{eqnarray}

So for $0\le i\le {2n-1},\quad 0\le j \le {2n-1}$ we have
\be \mbox{Res}_{w}\,(x^{i+j})=\left\{ \begin{array}{ll} 0&\mbox{if}\,\,
n\nmid (i+j+1),\\ 0&\mbox{if}\,\,(i+j+1)=n,\\
1/t&\mbox{if}\,\,(i+j+1)=2n,\\
-\,(2c)/t&\mbox{if}\,\,(i+j+1)=3n \end{array} \right. \ee
and the matrix $\eta _{ij}$ in the basis $\sqrt{t}\,\{1,x,\ldots ,
x^{2n-1}\}$ takes the form
\be \left( \begin{array}{cc} 0&J\\  J&{-2cJ} \end{array} \right) ,\ee
where $J$ is the $n\times n$ matrix
\be J= \left( \begin{array}{ccccc}0&0&\ldots &0&1\\
               0&0&\ldots &1&0\\
               \vdots &\vdots &\ldots &\vdots &\vdots\\
               0&1&\ldots &0&0\\
               1&0&\ldots&0&0\end{array} \right) . \ee

{\bf 4. Structure of Chiral Algebra.}
\medskip
Note first at all that since
\be x^{2n}+ 2c\,x^n- 1= \prod _{j=0}^{n-1} (x-a_j)(x-b_j),\ee
where $a_0,\ldots ,a_{n-1},b_0,\ldots ,b_{n-1}$ are distinct, we have
the {\bf C}-algebra isomorphism
\be {\cal R}= {\bf C}[x]/\left( x^{2n}+ 2cx^{n}- 1\right) \simeq
\bigoplus _{j=0}^{n-1}{\bf C}[x]/(x-a_j)\oplus \bigoplus _{j=0}^{n-1}
{\bf C}[x]/(x-b_j)\ee
(by the Chinese Reminder Theorem).

Let $\delta _j, \delta'_{j}\quad (j=0,\ldots ,n-1)$ be the elements
in ${\cal R}$
\begin{eqnarray}
\delta _j &\Longleftrightarrow &\left( 0,\ldots ,\frac{\bar x}{a_j},
0,\ldots ,0\right),\qquad \frac{\bar x}{a_j}\in{\bf C}[x]/(x-a_j),
\nonumber \\
&&\\
\delta'_{j}&\Longleftrightarrow &\left( 0,\ldots ,0,\ldots ,
\frac{\bar x}{b_j},\ldots ,0\right) ,\qquad \frac{\bar x}{b_j}\in{\bf C}[x]/
(x-b_j). \nonumber \end{eqnarray}

Then $\{\delta _0,\ldots ,\delta _{n-1},\delta'_{0},\ldots ,\delta'_{n-1}\}$
is a basis of ${\cal R}$, and its relation with the monomial basis
$\{1,x,\ldots ,x^{2n-1}\}$ is given by the formula
\[ \left( \begin{array}{c} 1\\ x\\ \vdots \\ x^{2n-1}\end{array}
\right) \]
\be =\left( \begin{array}{cccccccc} 1&1&\ldots &1&1&1&\ldots &1\\
a&a{\omega }&\ldots &{}&b\epsilon &b\epsilon \omega &\ldots &b\epsilon
\omega ^{n-1}\\
\vdots &\vdots &\ldots &\vdots &\vdots &\vdots &\ldots &\vdots \\
a^{2n-1}&a^{2n-1}\omega ^{2n-1}&\ldots &{}&b^{2n-1}\epsilon ^{2n-1}&{}
&\ldots &{}\end{array} \right) \left( \begin{array}{c}
\delta _0\\ \delta _1\\  \vdots \\ \delta _{n-1}\\ \delta '_0\\
\vdots \\ \delta '_{n-1}\end{array} \right) . \ee

Therefore the topological-topological coupling with respect to basis \break
$\{ \sqrt{t}\,(\delta _0,\ldots ,\delta _{n-1},\delta '_0,\ldots ,
\delta '_{n-1})\}$ is given by formula
\be \frac1{n(a^n + b^n)}\left( \begin{array}{cc}
a^{-(n-1)}\left( \begin{array}{cccc}1&0&\ldots &0\\
0&\omega &\ldots &{0}\\
\vdots &\vdots &\ddots &\vdots \\
0&0&\ldots &\omega ^{n-1}\end{array} \right) & 0\\
0&b^{-(n-1)}\epsilon \left( \begin{array}{cccc}1&0&\ldots &0\\
                          0&\omega &\ldots &0\\
                          \vdots &\vdots &\ddots &\vdots \\
                          0&0&\ldots &\omega ^{n-1}\end{array} \right)
\end{array} \right) . \ee

\newpage
{\bf 5. Lemma.}

The group of {\bf C}-algebra automorphisms of ${\cal R}$  is generated
by the transformation
\be \theta \colon x \rightarrow \omega x.\ee
Hence,
\be \theta \colon \left( \begin{array}{c} 1\\ x\\ \vdots \\
x^{2n-1} \end{array} \right) \rightarrow \left( \begin{array}{cc}
\left( \begin{array}{cccc} 1&0&{}&0\\ 0&\omega &{}&{}\\
\vdots &\vdots &\ddots &\vdots \\ 0&0&{}&\omega ^{n-1}\end{array} \right)
&0\\
0&\left( \begin{array}{cccc} 1&0&{}&0\\0&\omega &{}&0\\
 \vdots &\vdots &\ddots &\vdots\\0&0&{}&\omega ^{n-1}\end{array} \right)
 \end{array} \right) \left( \begin{array}{c} 1\\ x\\ \vdots \\ x^{2n-1}
\end{array} \right) ,\ee
\be \theta \colon \left( \begin{array}{c}\delta _0\\ \delta _1\\
 \vdots \\ \delta '_0\\ \vdots \\  \delta '_{n-1}\end{array} \right)
 \rightarrow \left( \begin{array}{cc} \left( \begin{array}{cccc}
 0&0&\ldots &1\\ 1&0&\ldots &0\\  \vdots &\ddots &\ddots &\vdots \\
                    0&\ldots &1&0\end{array} \right) &0\\
          0&\left( \begin{array}{cccc} 0&0&\ldots &1\\
                      1&0&\ldots &0\\
                      \vdots &\ddots &\ddots &\vdots\\
                      0&\ldots &1&0\end{array} \right) \end{array} \right)
\left( \begin{array}{c} \delta _0\\ \delta _1\\  \vdots \\
          \delta '_0\\ \vdots \\ \delta '_{n-1}\end{array} \right) .\ee
Let us go now to the consideration of topological-antitopological couplings
$(g_{i{\bar j}})$. As was shown in ref. [1] this matrix should satisfy
the equation
\be \bar \partial (g\,\partial g^{-1}) - \left[ C,gC^{+} g^{-1}\right]
= 0 \ee
and the reality constraint
\be \eta ^{-1}g(\eta ^{-1}g)^{\ast} = 1.\ee

Note first of all that in the ring ${\cal R}$ we have
\be w= t\left( \frac{x^{2n+1}}{2n+1}+ 2c\,\frac{x^{2n+1}}{n+1}- x\right)
=\frac{-2nt}{2n+1}\left( x-\frac{c}{n+1}\,x^{n+1}\right) .\ee
Then, for the multiplication up to $x$ in the basis $\{ \sqrt{t}, \sqrt{t}x,
\ldots ,\sqrt{t}x^{2n-1}\}$ there corresponds the matrix
\be \left( \begin{array}{ccccccc}0&1{}&{}&{}&{}&{}\\
{}&0&1&{}&{}&{}&{}\\
{}&{}&{}&\ddots &{}&{}&{}\\
{}&{}&{}&{}&1&{}&{}\\
{}&{}&{}&{}&{}&\ddots &{}\\
1&{}&{}&{}&(-2c)&\ldots &0\end{array} \right)  \ee
and for the multiplication up to $x^{n+1}$ there corresponds the matrix
\be \left( \begin{array}{cccccccc}{}&{}&{}&{}&0&1&{}&{}\\
{}&{}&{}&{}&{}&{}&\ddots &{}\\
{}&{}&{}&{}&{}&{}&{}&1\\
1&{}&{}&{}&(-2c)&{}&{}&{}0\\
0&1&{}&{}&0&{-2c}&{}&{}\\
{}&\ddots &\ddots &{}&{}&{}&{}&{}\\
{}&{}&{}&1&{}&{}&{}&{-2c}\\
(-2c)&{}&{}&0&(1+4c^2)&{}&{}&0\end{array} \right) , \ee
Hence the matrix $C$ in this basis has the form
\begin{eqnarray}
C &=& \left( \begin{array}{cccccccc} 0&1&{}&{}&{}&{}&{}&{}\\
{}&\ddots &\ddots &{}&{}&{}&{}&{}\\
{}&{}&0&1&{}&{}&{}&{}\\
{}&{}&{}&0&1&{}&{}&{}\\
{}&{}&{}&{}&0&1&{}&{}\\
{}&{}&{}&{}&{}&\ddots &\ddots &{}\\
1&{}&{}&{}&{(-2c)}&{}&{}&0\end{array} \right) \nonumber \\
&-& \frac{c}{n+1}\left( \begin{array}{cccccccc}
{}&{}&{}&{}&0&1&\ldots &0\\
{}&{}&{}&{}&{}&{}&\ddots &{}\\
1&{}&{}&{}&(-2c)&{}&{}&1\\
0&1&{}&{}&{}&(-2c)&{}&{}\\
{}&{}&\ddots &{}&\vdots &{}&\ddots &{}\\
{}&{}&{}&1&0&{}&{}&(-2c)\\
(-2c)&{}&{}&0&(1+4c^2)&{}&{}&0\end{array} \right) . \end{eqnarray}

As for the matrix $(g_{i{\bar j}})$ it has to be invariant relative to the
automorphism $\theta $ of ${\cal R}$ and hence it should have the form
\be \left( \begin{array}{cccccccc}
g_{0\bar 0}&0&\ldots &0&g_{0\bar n}&0&\ldots &0\\
0&\ddots &\ldots &\vdots &0&\ddots &\ldots &\vdots \\
\vdots &{}&\ddots &0&{}&{}&\ddots &0\\
0&{}&{}&g_{n-1,\overline {n-1}}&{}&{}&{}&g_{n-1,\overline {2n-1}}\\
{\bar g}_{0\bar n}&0&\ldots &0&\ddots &{}&{}&{}\\
0&\ddots &{}&{}&{}&\ddots &{}&{}\\
\vdots &\ddots &{}&{}&{}&\ddots &{}&{}\\
0&{}&{}&{\bar g}_{n-1,\overline {2n-1}}&{}&{}&{}&g_{2n-1,\overline {2n-1}}
\end{array} \right) . \ee

Let us consider now the reality constraint. For this purposes it is
convenient to use the basis ${\phi _j}$ in ${\cal R}$:
\be \sqrt{t}\,\{ 1,x,\ldots ,x^{n-1},(x^n+ c),\ldots ,(x^{2n-1}+cx^{n-1})\},
\ee
so that in this basis the matrix $\eta $ takes a very simple form
\be \eta \to \left( \begin{array}{cc} 0&J\\ J&0\end{array} \right) , \ee
where the matrix $J$ is given by the formula (25). From this we have
\be \eta =\eta ^{\ast} =\eta ^{-1}\ee
and from the reality constraint and $g^* =g^t$, it follows that
\be g\in SO(J,{\bf C}^{2n}), \ee
\be g= \exp \left\{ i\left( \begin{array}{cc} A&B\\ B^\ast &A^{\ast}
\end{array} \right) \right\} . \ee

Let us consider now the behavior as $c \to 0$. In the basis
$\{1,x,\ldots , x^{2n-1}\}$ we have
\be (g_{i\bar j})=(g_{i\bar j}^{(0)}) +c \left( \begin{array}{cccccc}
{}&{}&{}&h_1&{}&{}\\
{}&0&{}&{}&\ddots &{}\\
{}&{}&{}&{}&{}&h_n\\
\bar h_1&{}&{}&{}&{}&{}\\
{}&\ddots &{}&{}&{}&0\\
{}&{}&\bar h_n&{}&{}&{}\end{array} \right) ,\ee
\be
C=\left( \begin{array}{ccccc} 0&{}&1&{}&{}\\
{}&0&\ddots &{}&{}\\
{}&{}&{}&\ddots &1\\
1&{}&{}&{}&0\end{array} \right)  + cC'. \ee
So that in the limit $c \to 0$ we have
\be
w(x) \sim t\left( \frac{x^{n+1}}{n+1}- x\right) \ee
and we get the Toda equations.

It is also possible to get the precise results for arbitrary value
of it, if we use the theory of Chebyshev polynomials.

Let us remind that the Chebyshev polynomial $U_{k}(t)$ is defined
by the recurrent relation
$$ U_{k+1}(t)= 2t\,U_{k}(t) - U_{k-1}(t)\eqno(49)$$
and by "initial conditions" \be U_{0}(t)=
1,\quad U_{1}(t)= 2t.\ee So we have \be U_{k}(\cos \theta )=
\frac{\sin (k+1)\theta }{\sin \theta }\ee and the generating
function for these polynomials \be \frac1{1-2tz+z^2}= U_{0}(t)+
U_{1}(t)z+\cdots +U_{k}(t)z^{k}+\cdots ; \quad |z| < 1;\quad |t|<
1.\ee Substituting $z\to iz$, $t\to it$, we obtain \be
\frac1{1+2tz-z^2}= {\tilde U}_{0}(t)+{\tilde U}_{1}(t)z+\cdots +
{\tilde U}_{k}(t)z^{k}+\cdots , \ee where \be {\tilde U}_{k}(t)=
i^{k}U^{k}(it).\ee Then we have \be {\tilde U}_{k+1}(t)=
-2t\,{\tilde U}_{k}(t) + {\tilde U}_{k-1}(t),\ee \be {\tilde
U}_{0}(t)=1,\qquad {\tilde U}_{1}(t)= -2t,\ee and \be {\tilde
U}_{k}(-i\cos \theta )= i^{k}\,\frac{\sin (k+1)\theta } {\sin
\theta}.\ee Now it easy to prove the following

{\bf Lemma}. For $d\geq 0$
\be
\frac{x^{d+2}}{x^{2}+ 2tx -1}= {\tilde U}_{0}(t)\,x^{d}+ {\tilde U}_{1}(t)\,
x^{d-1}+\cdots +{\tilde U}_{d}(t)+ \frac{U_{d+1}(t)\,x+U_{d}(t)}
{x^{2}+ 2tx-1}.\ee

{\bf Proof}. Let us take the change $z\to x^{-1}$ in (52). We have
\be \frac{x^{2}}{x^{2}+2tx-1}={\tilde U}_{0}(t)+ {\tilde U}_{1}(t)\,
\frac1{x}+\cdots +{\tilde U}_{n}(t)\,\frac1{x^{n}}+\cdots ,\qquad
|x|>1.\ee
Hence,
\be \frac{x^{d+2}}{x^{2}+ 2tx-1}={\tilde U}_{0}(t)\,x^{d}+ {\tilde U}_{1}(t)\,
x^{d-1}+\cdots +{\tilde U}_{d}(t)+\frac{{\tilde U}_{d+1}(t)}{x}+
\frac{{\tilde U}_{d+2}(t)}{x^2}+\cdots \ee
and we have
\be \frac{x^{d+2}}{x^{2}+2tx-1}= {\tilde U}_{0}(t)\,x^d+ {\tilde U}_{1}(t)\,
x^{d-1}+ \cdots +{\tilde U}_{d}(t)+\frac{{\tilde U}_{d+1}(t)x+ C(t)}{x^{2}+
2tx-1}. \ee
If we put here $x=0$ we obtain $C(t)={\tilde U}_{d}(t)$.

To obtain the explicit expression for matrix $C$, note that this matrix is
determined by the multiplication of the element $\left( x-
\frac{c}{n+1}\,x^{n+1}\right) $ in the ring ${\cal R}$.

Hence,
\be C^{n}(1)= \left( x-\frac{c}{n+1}\,x^{n+1}\right) ^{n}= \left( 1-
\frac{c}{n+1}\,x^{n}\right) ^{n}x^{n}=
A_{n}(c)+ B_{n}(c)x^{n},\ee
\be C^{n}(x^{n})= B_{n}(c)+ (A_{n}(c)- 2cB_{n}(c))(x^{n}). \ee
So,
\be {\bf C}\cdot 1+ {\bf C}\cdot x^{n}\rightarrow ^{{ C}^{n}} {\bf C}
\cdot 1+ {\bf C}\cdot x^{n},\ee
\be \left( \begin{array}{c} {C}^{n}(1)\\
{C}^{n}(x^{n}) \end{array} \right) =\left( \begin{array}{cc}A&B\\
B&A-2cB\end{array} \right) \left( \begin{array}{c}1\\
x^{n} \end{array} \right) ,\qquad A= A_{n},\quad B=B_{n}.\ee
From here we obtain the characteristic equation
\be T^2 - 2(A-cB)T+ (A^2- 2cAB- B^2)= (T-\lambda )(T- \mu ),\ee
where
\begin{eqnarray}
\lambda &=& A- cB+ \sqrt {1+c^2}\,B= A- (c- \sqrt {1+c^2})B,\nonumber \\
&&\\
\mu &=& A- cB- \sqrt {1+c^2}\,B= A- (\sqrt {1+c^2}+ c)B.\nonumber
\end{eqnarray}
Let us define
\begin{eqnarray}
\lambda _{n}(t)&=& A_{n}(t)- (t-\sqrt {1+ t^2})\,B_{n}(t),\nonumber \\
&&\\
\mu _{n}(t)&=& A_{n}(t)- (t+ \sqrt {1+t^2})\,B_{n}(t).\nonumber\end{eqnarray}
We have
\be B_{n}(c) \neq 0\Leftrightarrow \lambda _{n}(c) \neq \mu _{n}(c).\ee
In this situation
\[ {\bf C}\cdot 1 + {\bf C}\cdot x^{n}={\bf C}\cdot \phi +{\bf C}\cdot
\phi ' \]
with
\begin{eqnarray}
\phi &=& 1+ \left( \frac{\lambda _{n}(c)}{B_{n}(c)} - \frac{A_{n}(c)}{B_{n}
(c)}\right) x^n,\nonumber \\
&&\\
\phi '&=& 1+ \left( \frac{\mu _{n}(c)}{B_{n}(c)}- \frac{A_{n}(c)}{B_{n}(c)}
\right) x^n. \nonumber \end{eqnarray}
Let us define
\be
\phi _{0}=\phi ,\quad \phi _1 =\frac{c\phi }{(\lambda _{n}(c))^{1/n}},
\ldots ,\phi _{j}=\frac{c^{j}\phi }{(\lambda _{n}(c))^{j/n}},\ldots,
\phi _{n-1}= \frac{c^{n-1}\phi }{(\lambda _{n}(c))^{(n-1)/n}},\ee
\be \phi '_{0}= \phi ',\quad \phi '_{1}= \frac{c\phi '}{(\mu _{n}(c))^{1/n}},
\ldots ,\phi '_{j}= \frac{c^{j}\phi '}{(\mu _{n}(c))^{j/n}},\ldots ,
\phi '_{n-1}= \frac{c^{n-1}\phi '}{(\mu _{n}(c))^{(n-1)/n}}.\ee

Then $\{ \phi _{j},\phi '_{j}\} _{j=0}^{n-1}$ is the basis of ${\cal R}$ and
 we have
\[ {C}\left( \begin{array}{c}\phi _{0}\\
\vdots \\
\phi _{n-1}\\
\phi '_{0}\\
\vdots \\
\phi '_{n-1}\end{array} \right) \]
\be =\left( \begin{array}{cc}(\lambda _{n}(c))^{1/n}\left( \begin{array}{cccc}
0&1&{}&{}\\
{}&0&\ddots &{}\\
{}&{}&\ddots &1\\
1&{}&{}&0 \end{array} \right) &0\\
0&(\mu _{n}(c))^{1/n}\left( \begin{array}{cccc}0&1&{}&{}\\
{}&0&\ddots &{}\\
{}&{}&\ddots &1\\
1&{}&{}&0\end{array} \right) \end{array} \right) \left( \begin{array}{c}
\phi _{0}\\ \vdots \\ \phi _{n-1}\\ \phi '_{0}\\ \vdots \\ \phi '_{n-1}
\end{array} \right) . \ee
It is more convenient to rewrite eq. (73) in another basis
\be
{C}\left( \begin{array}{c} \phi _0\\ \phi _0'\\ \phi _1\\ \phi _1'\\ \vdots \\
\phi _{n-1}\\ \phi _{n-1}'\end{array} \right) =\left( \begin{array}{ccccc}
0&D&{}&{}&{}\\
{}&0&D&{}&{}\\
{}&{}&\ddots &\ddots &{}\\
{}&{}&{}&\ddots &D\\
D&{}&{}&{}&0 \end{array}\right) \left( \begin{array}{c}\phi _0\\
\phi '_{0}\\
\phi _{1}\\
\phi '_{1}\\
\vdots \\
\phi _{n-1}\\
\phi '_{n-1} \end{array} \right) , \ee
where
\be D= \left( \begin{array}{cc}(\lambda _{n}(c))^{1/n}&0\\
0&(\mu _{n}(c))^{1/n}\end{array} \right) .\ee
In this basis the matrix of topological-antitopological couplings has the
form
\be \left( \begin{array}{cccc} G_{0}&{}&{}&{}\\
{}&G_{1}&{}&{}\\
{}&{}&\ddots &{}\\
{}&{}&{}&G_{n-1}\end{array} \right) \ee
and the basic equation (9) takes the form
\be {\bar \partial }(G_{j}\,\partial G_{j}^{-1})- G_{j+1}G_{j}^{-1}+
G_{j}\,G_{j-1}^{-1} = 0,\qquad j= 0,\ldots ,n-1;\quad G_{j}= G_{j+n}.\ee
So, this equation is the equation for the nonabelian Toda system, where the
quantities $G_{j}$ are elements of the non-Abelian group $SL(2,{\Bbb R})$.


\begin{thebibliography}{AAAA}

\bibitem{CV} S. Cecotti and C.Vafa, {\em Nucl. Phys.} {\bf B367}, 359 (1991)
\bibitem{Du1} B. Dubrovin, {\em Nucl. Phys.} {\bf B379}, 627 (1992)
\bibitem{Du2}  B. Dubrovin, {\em Geometry and integrability of
topological}\qquad\qquad\break
{\em -antitopological fusion}, Preprint INFN - 8/92 - DSF
\bibitem{Pe} A. Perelomov, {\em Phys. Lett.} {\bf B285}, 217 (1992)

\end{thebibliography}
\end{document}